\documentclass{aa}
\usepackage{txfonts}
\usepackage{graphicx}
\usepackage{natbib}
\usepackage{color}
\usepackage[hidelinks=true]{hyperref}
\definecolor{dark-red}{rgb}{0.4,0.15,0.15}
\definecolor{dark-blue}{rgb}{0.1,0.1,0.6}
\definecolor{light-blue}{rgb}{0.0,0.0,0.8}
\definecolor{medium-blue}{rgb}{0,0,0.5}
\hypersetup{
    colorlinks, linkcolor={dark-red},
    citecolor={light-blue}, urlcolor={medium-blue}
} 

\begin{document}

\title{Anomalous flows in a sunspot penumbra}

\author{Rohan E. Louis$^{1}$ \and Christian Beck$^{2}$ 
\and Shibu K. Mathew$^{3}$ \and P. Venkatakrishnan$^{3}$}

\institute{Leibniz-Institut f\"ur Astrophysik Potsdam (AIP),
	  An der Sternwarte 16, 14482 Potsdam, Germany \and
          National Solar Observatory, Sacramento Peak, 
          3010 Coronal Loop, Sunspot, New Mexico 88349, U.S.A. \and
          Udaipur Solar Observatory, Physical Research Laboratory,
          Dewali, Badi Road, Udaipur 313004, Rajasthan, India}

\date{Received ... 2012 / Accepted ...}

\abstract
   {The photospheric Evershed flow is a distinct characteristic of 
penumbrae that is closely associated with the photometric and magnetic 
structure of sunspots.}
   {We analyze the properties of an anomalous flow in the photosphere 
in a sunspot penumbra and compare it with those of the regular Evershed 
flow.}
   {High-resolution spectropolarimetric observations of active region 
NOAA 11271 were obtained with the spectro-polarimeter (SP) on board 
{\em Hinode}. We use the magnetic field properties derived by a 
Milne-Eddington inversion. In addition, we use Ca~{\sc II}~H images 
obtained by the broad-band filter instrument to study the low 
chromospheric response to the photospheric structure and 
Dopplergrams from the Helioseismic and Magnetic 
Imager to follow the evolution of the photospheric flows.}
   {We detect a blue-shifted feature that appeared on the limb-side 
penumbra of a sunspot and that was present intermittently 
during the next two hours. It exhibited a maximum blue-shift of 
1.6~km~s$^{-1}$, an area of 5.2~arcsec$^2$, and a maximum uninterrupted
lifetime of 1~hr. The blue-shifted feature, when present, lies parallel 
to red-shifts. Both blue and red shifts flank a highly inclined/horizontal 
magnetic structure that is radially oriented in the penumbra. The 
low-cadence SP maps reveal changes in size, radial position in the 
penumbra and line-of-sight (LOS) velocity of the blue-shifted feature, 
from one scan to the other. There was an increase of nearly 500~G 
in the field strength with the onset of the blue-shifts, particularly 
when the LOS velocity in the feature exceeded 1.5~km~s$^{-1}$. There 
was only a marginal reduction in the field inclination of about 10$^{\circ}$ 
with the increase in blue-shifts. In the chromosphere, intense, 
arc-shaped brightenings were observed close to the location of the 
photospheric blue-shifts, which extend from the edge of the umbral 
core to the penumbra-quiet Sun boundary. The intensity of these 
brightenings exceeds the background intensity by 30 to 65\% with 
the strongest and largest brightenings observed about 30~min 
after the strongest blue-shifts were detected at the photosphere. 
The close spatial proximity of the two phenomenon strongly 
suggests a causal relationship.}
    {The blue-shifted feature represents plasma motion that could 
be related to a magnetic structure that rises in the solar 
atmosphere and subsequently reconnects with the ambient 
chromospheric magnetic field of the sunspot or an inverse 
Evershed flow, which would be unique in the photosphere. This 
transient phenomena is presumably related to the dynamic stability 
of the sunspot because the corresponding umbral core separated 
two days later at the location of the blue-shifts and 
fragmented subsequently.}


\maketitle

\section{Introduction}
\label{intro}
The Evershed flow \citep[EF;][]{1909MNRAS..69..454E} is a 
characteristic property of sunspot penumbrae which has 
been studied extensively for several years \citep[][to mention a few]
{1987PASJ...39..329I,1992ApJ...398..359S,1994A&A...290..972R,
1996A&A...315..603B,1997Natur.389...47W,2003A&A...410..695M,
2004A&A...427..319B,2008A&A...480..825B}. The EF comprises a 
nearly horizontal and radial outflow of plasma. Consequently, 
the centre- and limb-side penumbra are blue- and red-shifted 
respectively, when the sunspot is located off the disk centre. 
In the upper solar atmosphere the direction of the EF reverses 
and flows into the sunspot and as such is referred to as the 
reverse/inverse EF \citep{1913ApJ....37..322S}. However, the 
formation of a penumbra is sometimes associated with flows that 
have an opposite sign as the EF in the photosphere 
\citep{2012ASPC..455...61S} and can even be supersonic 
\citep{2013A&A...552L...7L}. \citet{2008ApJ...676..698B} 
showed an example of an an oppositely directed EF 
associated with decaying penumbral filaments.

Penumbrae can also harbour the EF and a counter-EF at the 
photosphere, as in the case of $\delta$-spots where convergent 
flows slip past one another and return to the solar interior 
\citep{2002ApJ...575.1131L} or in regular sunspots which 
consist of disjoint penumbrae residing within the primary 
penumbra \citep{2013ApJ...770...74K}. Apart from the EF other 
anomalous flows are also known to exist in the penumbra. 
\citet{2010A&A...524A..20K} reported the existence of small 
downflow patches measuring 0\farcs5 in size that were 
sometimes associated with brightenings in the chromosphere. 
Since these patches were often seen on the centre-side 
penumbra and had the same polarity as the sunspot, their 
origin, according to \citet{2010A&A...524A..20K}, appeared 
to be unrelated to the EF. \citet{2011ApJ...727...49L} 
discovered large patches of supersonic downflows at/close 
to the umbra-penumbra boundary of sunspots with lifetimes 
of more than 14~hr that were accompanied by strong and 
long-lived chromospheric brightenings. 

The physical mechanism responsible for the EF is intimately 
connected to the photometric, magnetic and kinematic 
structure of the penumbra as well as its fine structure, 
which is in itself a highly debated topic 
\citep{1993ApJ...407..398T,1998ApJ...493L.121S,
2006ApJ...646..593R,2008ApJ...687..668B,2010ApJ...720.1417P,
2011ApJ...734L..18J,2013A&A...549L...4R,2013A&A...553A..63S}. 
The EF has been described in terms of a siphon flow 
\citep{1993ApJ...407..398T}, where the outflow is a result 
of the pressure gradient at the footpoints of a flux tube. 
On the other hand, the `moving-tube' model of 
\cite{1998ApJ...493L.121S} shows that the EF arises from a 
combination of hot plasma rising at the inner footpoint of 
the tube and a pressure difference from radiative losses. 
More recently, \citet{2011ApJ...729....5R} demonstrated that 
the Evershed flow in three-dimensional MHD simulations can be 
understood as the convective flow component in the direction 
of the magnetic field where the penumbral fine structure results 
from anisotropic magneto-convection. In this paper, we study 
the evolution of an atypical penumbral flow and compare its 
properties with the well-known photospheric EF.

\begin{figure}[!h]
\centerline{
\includegraphics[angle=90,width = \columnwidth]{gband-sunspot.eps}
}
\vspace{-10pt}
\centerline{
\hspace{-20pt}
\includegraphics[angle=90,width = \columnwidth]{horizontal-mag-field.eps}
}
\vspace{-17pt}
\caption{Leading sunspot in NOAA AR 11271. Top panels, clockwise 
from top left -- G-band intensity, LOS velocity, field 
inclination in the local reference frame and field strength. The 
MERLIN maps correspond to the last SP scan acquired from 
10:05--10:21~UT. Blue and red colours in the LOS velocity map 
correspond to blue- and red-shifts respectively. The 
arrow in the top left panel points to disc centre. Bottom 
panel: Arrows corresponding to the horizontal magnetic 
field overlaid on the LOS velocity map. The FOV under study 
is marked by the dashed rectangle in the top panels.}
\label{fig01}
\end{figure}

\section{Observations}
\label{data}
For this investigation, we utilize spectropolarimetric 
observations from the {\em Hinode} spectropolarimeter 
\citep[SP,][] {2001ASPC..236...33L,2008SoPh..249..233I} 
for active region (AR) NOAA 11271 observed on 2011 August 19. 
The AR was mapped by the SP in the fast mode from 8:05--10:21~UT 
resulting in a total of five scans that were 
acquired at 8:05--8:21~UT, 8:30--8:46~UT, 8:55--9:11~UT, 
9:35--9:51~UT and 10:05--10:21~UT, respectively. The SP 
recorded the four Stokes parameters of the Fe {\sc i} lines 
at 630~nm with a spectral sampling of 21.5~m\AA, a step 
width of 0\farcs29 and a spatial sampling of 0\farcs32 along 
the slit. The field-of-view (FOV) covered by the SP 
was 75\arcsec$\times$82\arcsec. The observations were reduced 
with the corresponding routines of the Solar-Soft 
package \citep{2013SoPh..283..601L} to yield Level-1 data. 
The AR was located at a heliocentric angle $\Theta$ of 
29$^\circ$. For this analysis, we used Level-2 maps 
comprising two-dimensional maps of the magnetic field 
strength, inclination, azimuth and line-of-sight (LOS) velocity 
that were obtained from inversions of the Stokes profiles 
with the MERLIN\footnote{Level 2 maps from MERLIN inversions 
are provided by the Community Spectro-polarimetric Analysis 
Center at the following link--http://www.csac.hao.ucar.edu/csac} 
\citep{2007MmSAI..78..148L} code. The inclination and azimuth 
were subsequently transformed to the local reference frame. We 
use the mean umbral velocity as the zero-velocity reference. 
All the SP maps were co-aligned with respect to the first 
scan. The SP observations were complemented with co-temporal 
broad-band filtergrams in Ca {\sc ii} H and G-band which 
had a spatial sampling of 0\farcs1 and a cadence of 2 min. 

We also utilized data from the Helioseismic 
and Magnetic Imager \citep[HMI;][]{2012SoPh..275..229S}, 
namely, continuum intensity filtergrams with a spatial 
sampling of 0\farcs5 and a cadence of 12~min to trace the 
evolution prior and after the SP observations and Dopplergrams 
at a cadence of 3~min to follow the evolution of the LOS velocity 
in between the SP maps.

\section{Results}
\label{result}

\begin{figure*}[!ht]
\centerline{
\includegraphics[angle=90,width = 0.98\textwidth]{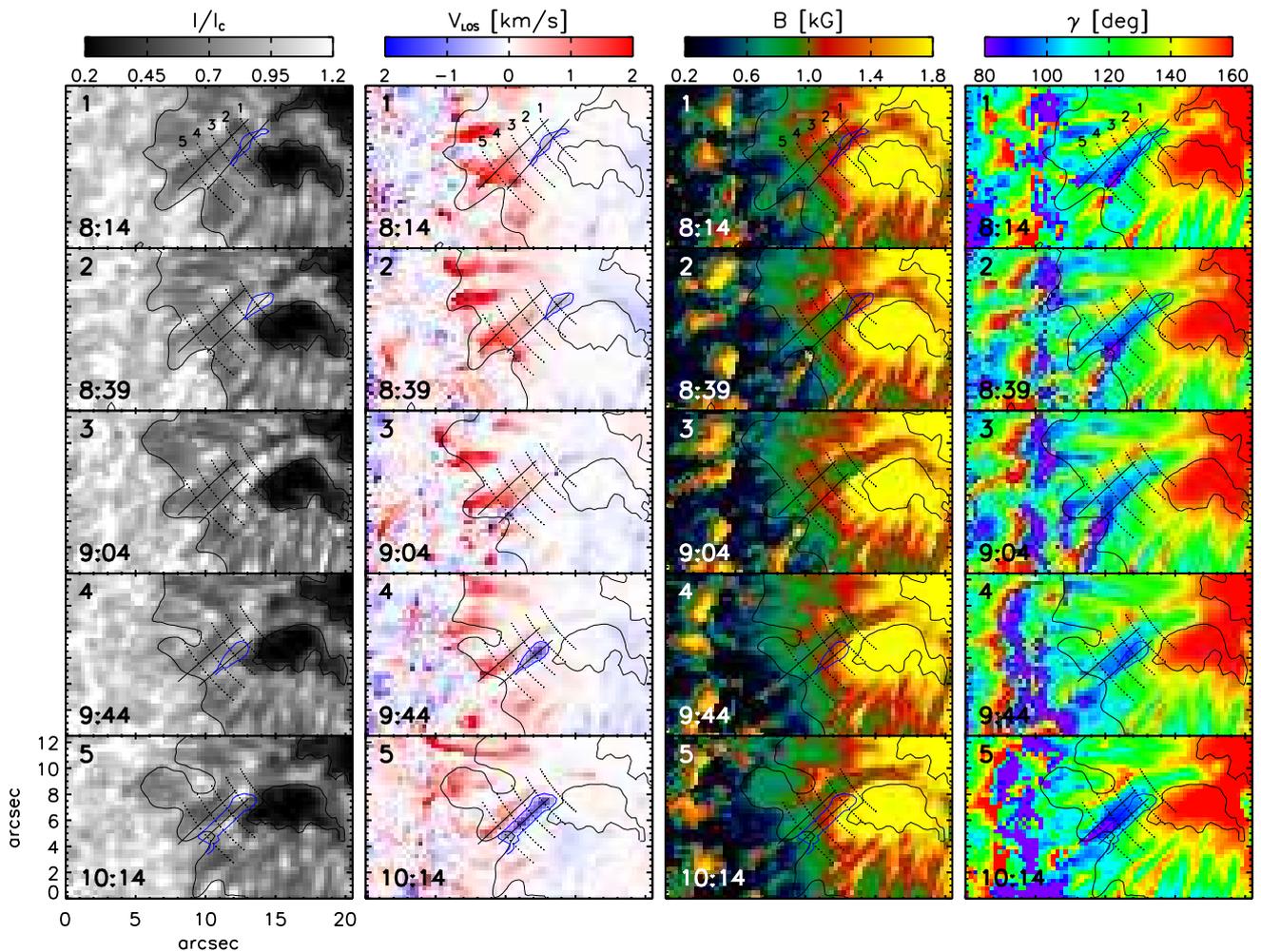}
}
\vspace{-15pt}
\caption{Temporal evolution of physical parameters. From left 
to right--G-band intensity, LOS velocity, field strength 
and inclination, respectively. The time indicated at the 
bottom left corner corresponds to the instant when the SP 
slit was over the BSF. The thin blue contour outlines the 
BSF and corresponds to $-0.2$~km~s$^{-1}$. Panel numbers 
are shown on the top left corner. Five azimuthal cuts at 
different radial distances and indexed 1--5 are indicated 
in panel 1. The plot also shows two radial lines passing 
through the BSF and RSF.}
\label{fig02}
\end{figure*}

\begin{figure*}[!ht]
\centerline{
\includegraphics[angle=90,width = 0.95\textwidth]{radial-variation.eps}
}
\vspace{-20pt}
\caption{Variation of parameters along the two radial lines 
passing through the blue- and red-shifted features in 
Fig.~\ref{fig02}. Time increases from left to right as 
indicated by the number at the bottom right corner, 
while the {\em blue crosses} and {\em red circles} 
correspond to the radial cuts along the blue- and 
red-shifted features, respectively. The left-most 
point on the $x$-axis corresponds to the upper end 
of the radial cut.}
\label{fig02a}
\end{figure*}

\subsection{Blue-shifted feature on the limb-side penumbra}
\label{blue}
The top left panel of Fig.~\ref{fig01} depicts the leading 
sunspot in AR 11271 which comprises two prominent 
light bridges (LBs) that separate the primary umbra into 
three umbral cores. The MERLIN maps shown in the figure 
correspond to the last SP scan acquired between 10:05--10:21~UT. 
The dashed rectangle in the panels represents our area of interest 
which clearly shows a blue-shifted filamentary structure on the 
limb-side penumbra that lies adjacent and parallel to a 
red-shifted feature (RSF). The blue-shifted feature (BSF) has 
a maximum blue-shift of 1.6~km~s$^{-1}$ and an area of 
5.2~arcsec$^2$. The BSF has a width and length of 1\arcsec and 
5\arcsec, respectively. The G-band image shows a bright 
penumbral grain close to the BSF and RSF in the inner penumbra. 
Both the BSF and RSF extend from the smaller umbral core to the 
penumbra-quiet Sun (QS) boundary. The BSF is located nearly 
perpendicular to the line-of-symmetry of the sunspot. In comparison, 
the LOS velocity in the RSF is about 1~km~s$^{-1}$ while the 
red-shifts in the limb-side penumbra are about 1.5~km~s$^{-1}$. The 
inclination map reveals that the BSF and RSF flank a horizontal 
elongated structure with the typical inclination in the BSF 
varying between 95--100$^\circ$. The RSF has a similar 
inclination as the BSF, but a few pixels close to the 
penumbra-QS boundary exhibit an opposite polarity. The field 
strength in the BSF varies from 1500--700 G, decreases with 
radial distance and is stronger than in the RSF by about 500~G. 
The bottom panel of Fig.~\ref{fig01} shows the arrows of the 
horizontal magnetic field overlaid on the LOS velocity map. The 
sunspot is of negative polarity and assuming a smooth azimuth 
around the spot, we find that the horizontal field is well 
aligned along the BSF and RSF. As one approaches the smallest 
umbral core close to the BSF, the horizontal magnetic field is 
more oblique to the two features. The figure also indicates an 
inward incursion of the penumbra-QS boundary at the BSF relative 
to the RSF.

\begin{figure}[!h]
\centerline{
\includegraphics[angle=0,width = \columnwidth]{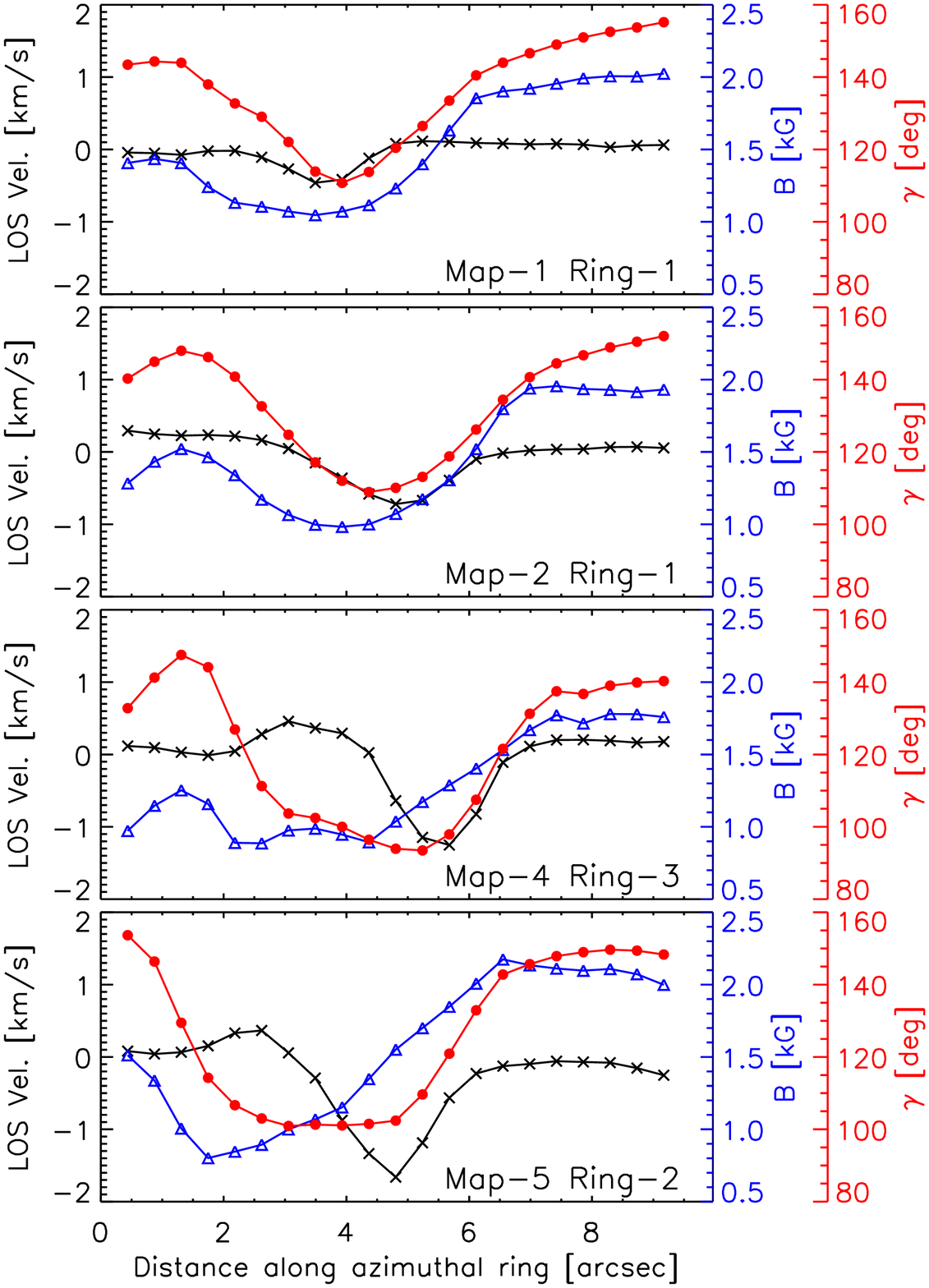}
}
\vspace{-10pt}
\caption{Variation of parameters along selected azimuthal cuts 
shown in Fig.~\ref{fig02} for different SP maps. The black line 
with {\em crosses} corresponds to the LOS velocity depicted on 
the left $y$-axis while field and strength and inclination are 
shown in {\em blue triangles} and {\em red circles}, respectively, 
and indicated on the right $y$-axis. The left-most pixel on the $x$-axis 
corresponds to the top of the azimuthal cut.}
\label{fig02b}
\end{figure}

\subsection{Temporal evolution of blue-shifted feature}
\label{temp}
The temporal evolution of the BSF and the related physical 
parameters in the SP maps are shown in 
Fig.~\ref{fig02} while the corresponding 
HMI LOS velocity maps are shown in Fig.~\ref{fig02c} below.
Figure~\ref{fig02} shows the maps from the five SP scans 
stacked in time from top to bottom. It is evident that 
the blue-shifts at the limb-side penumbra and enclosed 
in a thin blue contour evolve substantially over a period 
of 2~hr. The LOS velocity map corresponding to panel 3 
is nearly devoid of any discernible blue-shifts, with 
the exception of extremely weak blue-shifts close to 
the small umbral core and at the outer penumbral 
boundary. We defined the central axes of the BSF and 
RSF in each scan manually by straight lines. These 
radial lines indicate that the red-shifts are present, 
either behind or alongside the BSF, when the latter is 
observed but exist even in the latter's absence as 
indicated in panel 3. Values along the central axes 
are shown in Fig~\ref{fig02a}. Additionally, we 
laid a series of five azimuthal rings, each radially 
separated by 1.7\arcsec, at a fixed position through 
the corresponding section of the penumbra. The 
variation of the various parameters along a few 
selected azimuthal cuts are shown in Fig.~\ref{fig02b}. 
These cuts pass through pixels where the LOS velocity 
in the BSF was maximum.

Panels 4 and 5 of Fig.~\ref{fig02} show that the end 
of the blue-shifted blob close to the umbra tends to move 
closer to the inner penumbra with time. A similar 
feature is also seen in panels 1 and 2 although the 
blue-shifts are weaker than those seen nearly 1~hr 
later. The maximum blue-shifts observed from the scans 
indicated in panels 1, 2, 4 and 5 are 0.5~km~s$^{-1}$, 
0.8~km~s$^{-1}$, 1.4~km~s$^{-1}$ and 1.6~km~s$^{-1}$, 
respectively. The figure also shows that the radial width 
of the BSF evolves with time, reaching its maximum during 
the last SP scan/map. Furthermore, the position of the 
BSF is closer to the larger umbral core in panels 1 and 2, 
while it lies further out in the penumbra about 1 hr later 
(panels 4 and 5). As stated in Sect.~\ref{blue} and 
seen in the G-band panels of Fig.~\ref{fig02}, the 
penumbra-QS boundary is displaced inwards along the 
radial line passing through the BSF. A comparison of 
the field strengths in the red and blue-shifted areas 
reveals that the latter is stronger by $\approx$500 G. 
The maps of the field inclination indicate that the 
blue-shifts lie on the edge of a highly inclined or 
even horizontal magnetic structure that persists 
throughout the period of the SP scans. In panels 
1 and 5 the polarity in this filamentary structure 
close to the penumbra-QS boundary has an opposite 
sign as the sunspot. The mean field strength and 
inclination in the BSF over the two hour duration 
is about 1100~G and 105$^\circ$, respectively.

Figure~\ref{fig02a} shows that the BSF is brighter 
than the RSF by nearly 10--15\% for most of its 
radial length. There is an increase in the field 
strength by about 400~G along the length of the 
BSF only when the LOS velocity is stronger than 
1~km~s$^{-1}$, as is the case in maps 4 and 5. In 
comparison, the field strengths in maps 1 and 2 
are about 1100~G, when the LOS velocity is less 
than 1~km~s$^{-1}$. The field inclination, on the 
other hand, reduces by about 10$^{\circ}$ with the 
increase in blue-shifts.

Figure~\ref{fig02b} shows that the field strength is 
weaker on the side of the RSF than it is on the side 
of the BSF. It is also observed that when the blue-shifts 
increase with time, as seen between maps 2 and 4, the 
increase in the field strength is around 300--500~G in the 
BSF. With the exception of the first SP map, the position 
of the minimum field strength is seen at or near the RSF 
with values of about 1000~G. As stated earlier, the BSF 
and RSF appear on the edge or side of a nearly horizontal 
magnetic structure, with the minimum inclination varying 
between 95$^{\circ}$--110$^{\circ}$. On either side of 
this horizontal magnetic structure, the field is closer 
to a vertical orientation by 30$^{\circ}$.

\subsection{Recurrence of blue-shifts}
\label{recur}
\begin{figure}[!h]
\centerline{
\includegraphics[angle=0,width = \columnwidth]{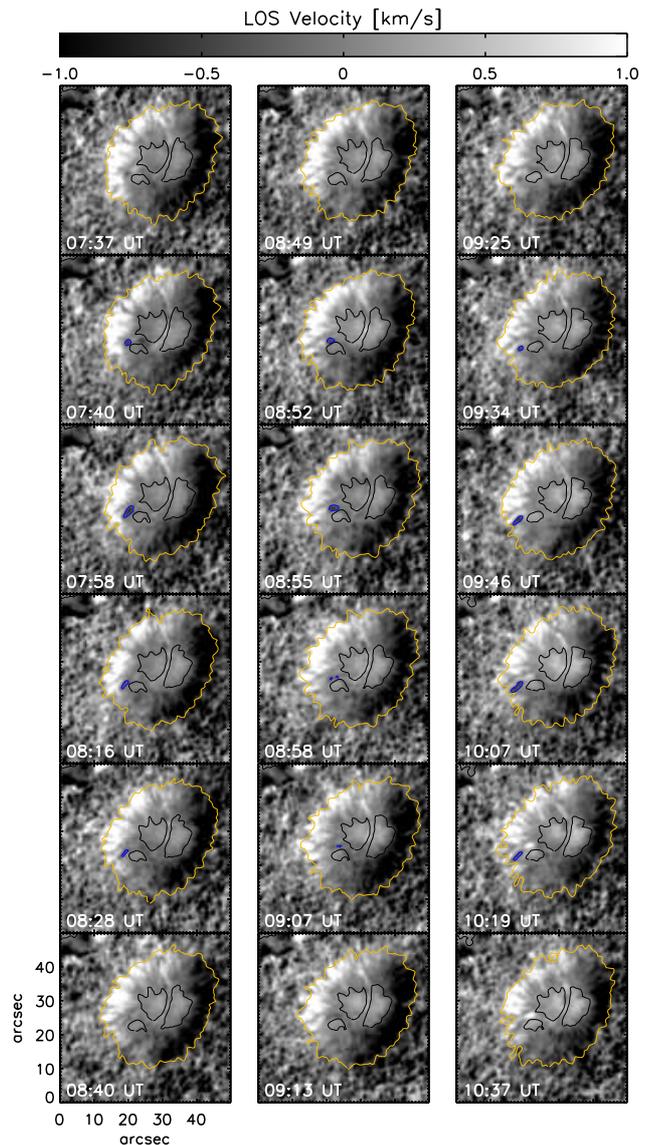}
}
\vspace{-15pt}
\caption{Evolution of photospheric blue-shifts seen from 
a time sequence of HMI Dopplergrams. The panels correspond
to three events that were detected during the time of the 
{\em Hinode} SP scans. The {\em blue} contour outlines the BSF.}
\label{fig02c}
\end{figure}

The BSF can also be identified in the time sequence of 
HMI Dopplergrams (blue contours in Fig.~\ref{fig02c}). 
The HMI data reveal that the BSF in the SP maps 
corresponds to three separate appearances of such 
features next to the smallest umbral core 
in the HMI Dopplergrams. The first event starts at 
7:40~UT and persists for nearly 1~hr 
(left column of Fig.~\ref{fig02c}) while the second 
event starts around 8:52~UT but only survives for 
about 20~min. The third event begins at 9:28~UT and 
has a similar lifetime as the first. It coincides 
with the duration of the fourth and fifth SP scans. 
We also find a couple of events during the early part 
of August 19 but these are short-lived events with 
lifetimes of less than 20~min. Similar BSFs are also 
observed on August 20 when the smaller umbral core 
started to separate from the sunspot. The HMI 
Dopplergrams give the impression that the BSFs 
evolve in-situ within the area close to the smaller 
umbral core. We determined the mean LOS velocity in 
the BSF using contours (Fig.~\ref{fig02c}) enclosing 
blue-shifts stronger than 100~m~s$^{-1}$. This was 
done for the time sequence of HMI Dopplergrams for 
a comparison with the evolution of the intensity 
in the low chromosphere described in the next section.

\subsection{Chromospheric response}
\label{chrom}
\begin{figure}[!h]
\centerline{
\hspace{20pt}
\includegraphics[angle=90,width = 1.1\columnwidth]{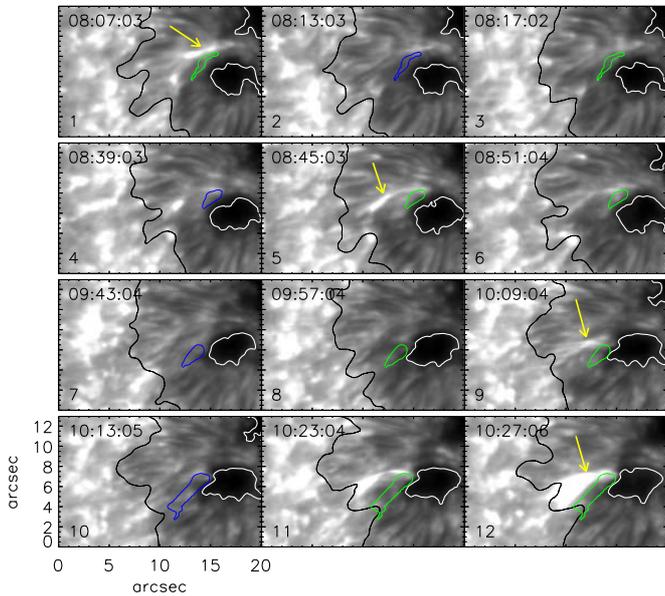}
}
\vspace{-10pt}
\caption{Chromospheric response to photospheric blue-shifts. 
The four rows correspond to SP scans 1, 2, 4 and 5, respectively. 
The blue contour represents the BSF at the time when the SP 
slit was above the feature while the green contour represents 
the same at other instances of time. See text for description 
of yellow arrows. The white and black contours correspond to 
the umbral and penumbral areas, respectively, determined from 
the G-band intensity.}
\label{fig03}
\end{figure}

In this section we analyze the influence of the 
photospheric blue-shifts on the overlying chromosphere. 
Figure~\ref{fig03} shows a selected set of Ca~{\sc ii}~H 
filtergrams acquired close to the time interval when the 
SP slit was over the blue-shifted patch. It is evident 
that the appearance of the BSF in the photosphere 
affects the chromosphere. The effect is seen in the form
of intense arc-shaped brightenings nearly 9\arcsec~in 
length, that extend from the small umbral core to the 
penumbra-QS boundary (yellow arrows). One end of this 
arc-shaped brightening in the inner/mid penumbra is 
co-spatial or very close to the BSF. In some cases, 
there can be isolated bright patches (panel 4) in 
the chromosphere that later evolve into the arc-shaped 
structure described above. The brightening is remarkably 
strong and conspicuous during the last SP map where 
the blue-shifts were also the strongest (panel 12).

\begin{figure}[!h]
\centerline{
\includegraphics[angle=90,width = \columnwidth]{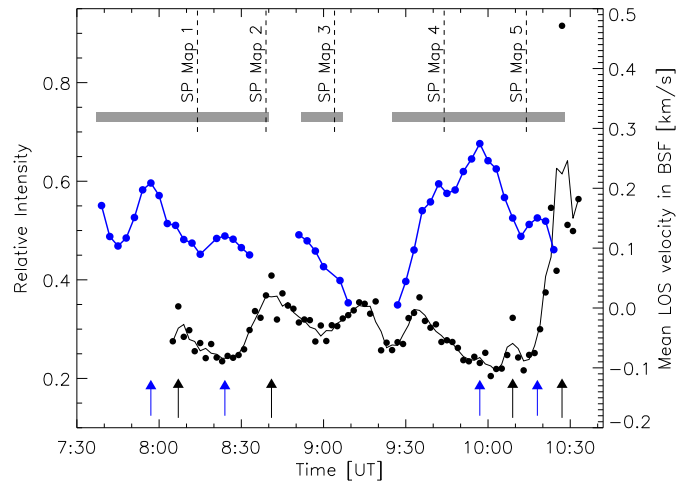}
}
\vspace{-5pt}
\caption{Chromospheric light curve in the vicinity of the 
photospheric blue-shifts. The solid black line represents 
a 3-point smoothing to the observed relative intensity 
({\em black filled circles}). The relative intensity is 
the excess over the time averaged background intensity. 
The grey horizontal bars represent the 
moments of the BSF's existence. The blue {\em filled circles} 
and solid line correspond to the mean LOS velocity in the
BSF estimated from HMI Dopplergrams and whose scaling is 
indicated on the right $y$-axis. The black and blue arrows 
at the bottom half of the plot indicate local maxima in the 
chromospheric light curve and photospheric LOS velocities, 
respectively.}
\label{fig04}
\end{figure}

We derived a light curve for the region, where the 
strong brightenings were observed, from the time 
sequence of the measured relative intensity 
variation. All pixels inside a hand-drawn contour, 
whose intensity exceeded the time-averaged 
intensity inside the same region by 15\% or more, 
were selected. Their mean intensity as a function 
of time is shown in Fig~\ref{fig04}. The plot 
displays the relative intensity with respect 
to the background and shows that the excess 
intensity of the chromospheric brightening is 
between 30--65\%. The chromospheric light curve 
exhibits intermittent peaks during the course 
of the five SP scans, e.g. at 
8:07~UT, 8:41~UT, and 10:27~UT. The intensity 
peaks occur during the lifetime of the individual 
blue-shifted events that were identified in the 
HMI Dopplergrams. Furthermore, there appears to 
be a variable time-lag of 10 to 20~min between 
the evolution of the photospheric blue-shifts 
and the chromospheric brightenings (blue and black 
arrows in Fig.~\ref{fig04}). It is also evident 
from Fig.~\ref{fig04} that the largest blue-shift 
occurs about 30~min prior to the strongest 
chromospheric brightening. We note that the 
two peaks in the chromospheric light curve 
between SP maps 3 and 4 in Fig.~\ref{fig04} are 
due to brightenings at the edge of the mask 
which covers the penumbra-QS boundary and do 
not exhibit the characteristic arc-shaped 
structure seen in other filtergrams. 

\subsection{Temporal evolution of sunspot}
\label{evol}
\begin{figure*}[!ht]
\centerline{
\includegraphics[angle=90,width = 1\textwidth]{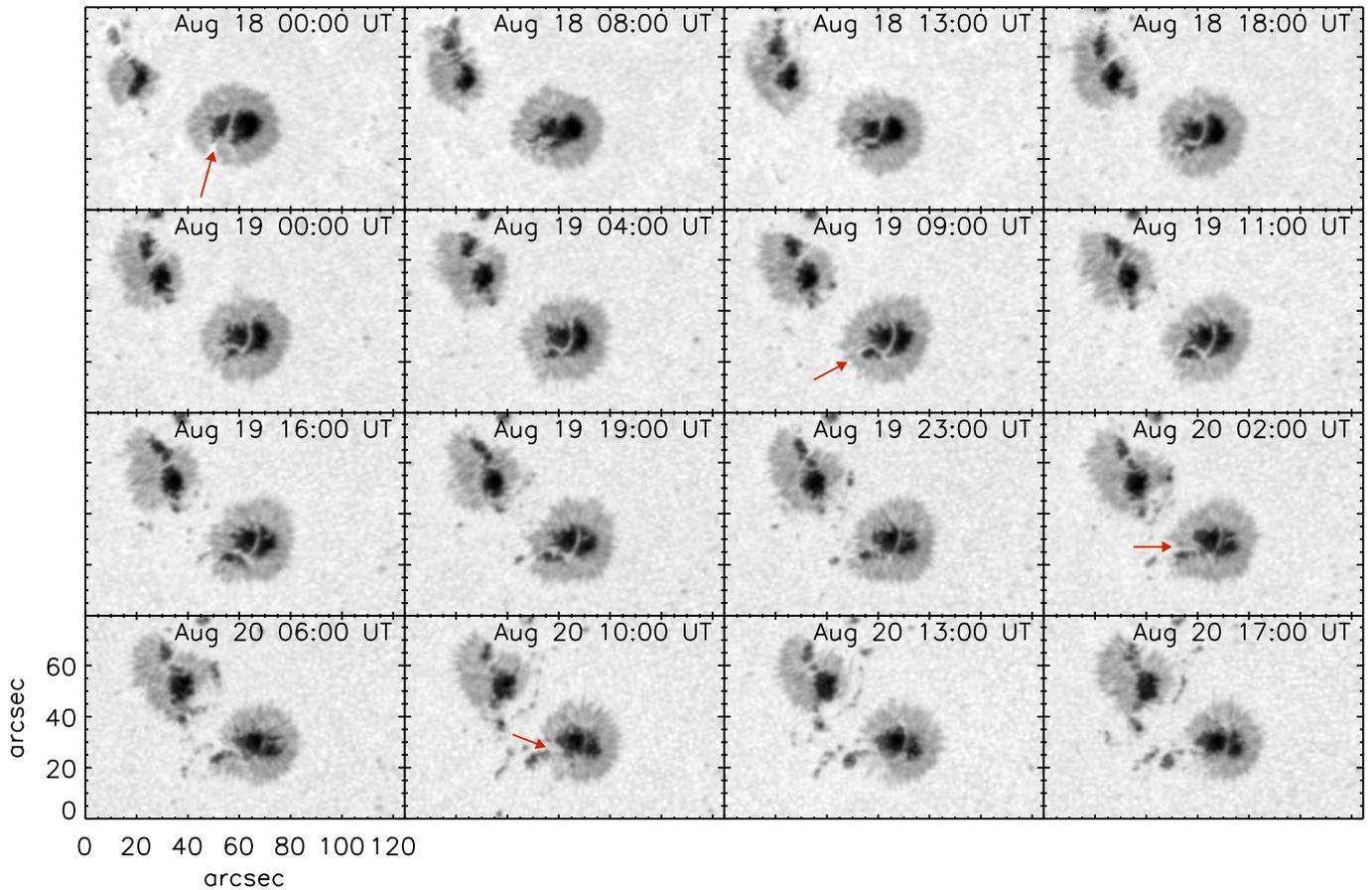}
}
\vspace{-5pt}
\caption{HMI continuum intensity filtergrams depicting 
the evolution of the leading sunspot in NOAA AR 11271 
over a period of  three days starting from 2011 
August 18. Solar north and west point to the top 
and right respectively.}
\label{fig05}
\end{figure*}

In order to relate the above observations with the 
global sunspot structure, we analyze the evolution 
of the latter in low-cadence HMI 
continuum intensity filtergrams spanning a duration 
of 3~days beginning on August 18. Figure~\ref{fig05} 
shows that the leading sunspot has a light bridge 
separating the umbra into two nearly equal halves 
on August 18. In addition, the azimuthal arrangement 
of the penumbra is disrupted at the location where
we observed the blue-shifts, nearly 32 hrs later. At 
this location, the penumbra is apparently absent, 
with granulation extending all the way up to the 
umbral core. However, the formation of penumbra 
proceeds rapidly and close to the end of August 18, 
the sunspot has a symmetric penumbra, with an 
additional small light bridge. At the same time, 
we see that the following spot increases in size. 
While the {\em Hinode} SP observations were made 
between 8:00-10:00~UT on August 19, first signs
of decay in the leading sunspot appear just 6~hr 
later, with small penumbral fragments separating 
from the sunspot. This decay process continues into 
the early part of August 20 wherein the entire 
section of the penumbra close to the smaller light
bridge starts to break away. This part of the sunspot 
eventually separates from the parent spot by 17:00~UT. 
The parent sunspot also partially recovers its penumbra 
at the location of fragmentation close to the end
of August 20.  

\section{Summary}
\label{summary}
We have analyzed the properties of a blue-shifted feature 
that appeared on the limb-side penumbra of a sunspot, using 
high resolution spectropolarimetric observations from 
{\em Hinode}. This feature was observed in four out of the 
five SP scans over a duration of 2 hr and exhibited a maximum 
blue-shift of 1.6~km~s$^{-1}$ and an area of 5.2~arcsec$^2$. 
The HMI Dopplergrams reveal that the blue 
shifts in the SP data were caused by the intermittent 
appearance of three individual, separate blue-shifted 
patches around the same location during the {\em Hinode} 
observing sequence. The lifetime of two of these events 
was nearly 1~hr while the third was short-lived and only 
persisted for about 20 min. Parallel and adjacent to the 
blue-shifted feature are red-shifted features which correspond 
to the regular Evershed flow. The blue-shifted feature and the 
Evershed red-shifts flank a highly inclined or nearly horizontal 
magnetic structure that is radially oriented in the penumbra. The 
mean field strength and inclination in the blue-shifted feature 
over the two hour duration are about 1100~G and 105$^\circ$, 
respectively. The SP maps represent a snapshot of the 
blue-shifted feature at different epochs as seen from the 
changes in its size, radial position in the penumbra and 
LOS velocity. There was an increase of nearly 500~G in the 
field strength with the onset of the blue-shifts, particularly 
when the LOS velocity in the feature exceeded 1.5~km~s$^{-1}$. 
There was only a marginal reduction in the field inclination 
of about 10$^{\circ}$ with the increase in blue-shifts.

In the chromosphere, intense, arc-shaped brightenings 
were observed close to the location of the photospheric 
blue-shifts, which extend from the edge of the smaller 
umbral core to the penumbra-QS boundary. The intensity 
of these brightenings exceed the background intensity by 
30 to 65\% with the strongest and largest brightenings 
observed during the largest blue-shifts. The chromospheric 
intensity in the neighbourhood of the blue-shifts exhibits 
intermittent peaks during the observing sequence. The 
strongest chromospheric brightening was observed about 
15~min later to the strongest blue-shift detected by 
{\em Hinode} and 30 min after the largest 
blue-shift seen in the HMI data. The other chromospheric 
brightenings occur about 10--20~min after the photospheric 
blue-shifts reach a maximum value as estimated from HMI 
Dopplergrams. We speculate that the close spatial and 
temporal proximity of the two phenomenon is suggestive 
of a causal relationship. 

\section{Discussion}
\label{discuss}
In the limb-side penumbra one would normally expect 
only red-shifts due to the radial outflow of plasma 
in the photosphere. The blue-shifts detected in the 
LOS velocity map represent plasma motion towards the 
observer. One possible scenario is that this plasma 
motion transports magnetic fields upwards that 
eventually interact with the overlying sunspot 
magnetic field. This could lead to reconnection, thereby 
producing the intense arc-shaped brightenings that 
occur in close spatial and temporal proximity. From 
the observed time-lag of 10--20~min, magnetic flux 
moving with speeds of about 1~km~s$^{-1}$ would reach 
low chromospheric heights of 600--1200 km, which would 
support the above idea. We also do not discard the 
possibility that the blue-shifts could be produced 
by a radial inflow of plasma. The LOS velocity from 
a sonic radial inflow of 7~km~s$^{-1}$ that makes an 
angle ($\phi$) of about 65$^\circ$ 
with the line-of-symmetry is given by 
$v\cdot\sin{\Theta}\cos{\phi}$. This yields a value of 
1.4~km~s$^{-1}$ for $\Theta=29^\circ$, which is in good 
agreement with our observations. This scenario would imply 
an inverse EF in the photosphere instead of the chromosphere, 
which would be unique. 

Additionally, the long term evolution of the sunspot seen 
from the HMI intensity filtergrams provides an additional 
clue that the region was dynamically unstable. Using 
numerical simulations, \citet{2011ApJ...731..108B} have 
analyzed the influence of non-linear magnetoconvection 
around magnetic flux tubes. Their simulations involve 
solving the nonlinear resistive magnetohydrodynamic 
equations in a 3D cylindrical domain, for a central 
cylindrical flux tube surrounded by an annular convection 
cell. They demonstrate that as convection evolves the 
annular cell breaks up in to smaller cells in the 
azimuthal direction which allows magnetic flux to slip 
between these cells away from the central tube through 
the process of flux erosion. Such a process reduces the 
magnetic pressure inside the flux tube leading to the 
growth of convection inside it. The stability of the 
central tube depends on the convection around it. The 
simulations also show that magnetic flux can be added 
to the central tube when flux caught in the surrounding 
convection is pushed towards it. Such a non-linear 
interaction between magnetic flux and the surrounding 
convection could have led to the formation of the 
penumbral section at the location of the blue-shifts 
in the latter half of August 18. The blue-shifts we 
have described could be the result of plasma moving 
towards the umbral core as a result of magnetic flux 
being pushed and added to the sunspot by the surrounding 
convective flows. The sunspot also exhibits a weak clockwise 
rotation, as seen from the HMI images (arrows in Fig.~\ref{fig05}), 
that could be associated with the buoyant rise of a magnetic 
flux rope \citep{2003SoPh..216...79B,2012ApJ...745...37F}. 
Such a rotation, albeit small, could play a destabilizing 
role, in addition to the surrounding convection, and would 
eventually sever the temporary penumbral attachment of this 
section of the sunspot. The onset of convective instabilities 
in rotating cylindrical flux tubes has been described in 
numerical experiments of \citet{2008MNRAS.387.1445B}. 
The location of the blue-shifted feature coincides well 
with the line along which the sunspot starts to split one 
day later while the chromospheric reaction associated with 
the blue-shifts is suggestive of reconnection events. The 
appearance of the blue-shifted feature thus likely was the 
indicator for the development of an instability in the 
magnetic field configuration at this place.

Our results necessitate the need for high resolution 
spectropolarimetric observations with good temporal 
resolution and coverage both in the photosphere and 
chromosphere so as to determine the physical mechanism 
responsible for such anomalous flows. Instruments such 
as the Blue Imaging Solar Spectrometer \citep[BLISS,][]
{2013OptEn..52h1606P} that has been planned at the 
1.5 m GREGOR telescope \citep{2012AN....333..796S}, the 
Interferometric BIdimensional Spectrometer \citep[IBIS;][]
{2006SoPh..236..415C} and the CRisp Imaging 
SpectroPolarimeter \citep[CRISP;][]{2008ApJ...689L..69S} 
will be crucial for such investigations. 

\section{Conclusions}
\label{conclu}
We find an anomalous flow pattern opposite to the regular 
Evershed flow, i.e. a strong blue-shift, in the limb-side 
penumbra of a sunspot at photospheric heights. Intermittent, 
arch-shaped brightenings appear in the chromosphere above 
the feature. The sunspot splits open roughly along the 
location of the observed blue-shift while ejecting one 
of its umbral cores. We conclude that the anomalous flow 
in our case very likely indicates a substantial change in 
the magnetic field topology of the sunspot and is a 
precursor of the later loss of magnetic flux. Anomalous 
flow patterns could play a crucial role as indicators for 
the dissolution and fragmentation of sunspots in their 
late life phase.

\begin{acknowledgements}
Hinode is a Japanese mission developed and launched by ISAS/JAXA, 
collaborating with NAOJ as a domestic partner, NASA and STFC (UK) 
as international partners. Scientific operation of the Hinode 
mission is conducted by the Hinode science team organized at 
ISAS/JAXA. This team mainly consists of scientists from 
institutes in the partner countries. Support for the post-launch 
operation is provided by JAXA and NAOJ (Japan), STFC (U.K.), NASA, 
ESA, and NSC (Norway). Hinode SOT/SP Inversions were conducted at 
NCAR under the framework of the Community Spectro-polarimtetric 
Analysis Center (CSAC; http://www.csac.hao.ucar.edu/). HMI data 
are courtesy of NASA/SDO and the HMI science team. They are provided 
by the Joint Science Operations Center -- Science Data Processing 
at Stanford University. R.E.L is grateful for the financial 
assistance from the German Science Foundation (DFG) under grant 
DE 787/3-1. We thank the referee for his helpful suggestions and 
comments.
\end{acknowledgements}

%\bibliographystyle{aa}
%\bibliography{louis_reference}

\end{document}